\documentclass[numbook]{svjour3}
\usepackage[numbers]{natbib}
\usepackage{bm,amssymb,amsmath,graphicx,color}
\begin{document}
\title{A brief history of simple invariant solutions in turbulence}
\author{Lennaert van Veen}
\institute{Ontario Tech University, Oshawa, Ontario, Canada. \email{lennaert.vanveen@uoit.ca}}
\begin{abstract}
When studying fluid mechanics in terms of instability, bifurcation and invariant solutions one quickly finds out how little can be done by pen and paper.
For flows on sufficiently simple domains and under sufficiently simple boundary conditions, one may be able to predict the parameter values at which 
the base flow becomes unstable and the basic properties of the secondary flow. On more complicated domains and under more realistic boundary 
conditions, such questions can usually only be addressed by numerical means. Moreover, for a wide class of elementary parallel shear flows the base flow remains stable
in the presence of sustained turbulent motion. In such flows, secondary solutions often appear with finite amplitude and completely unconnected to the base flow.
Only using techniques from computational dynamical systems can such behaviour be explained. Many of these techniques, such as for the detection and classification
of bifurcations and for the continuation in parameters of equilibria and time-periodic solutions, were developed in the late 1970s for dynamical systems with 
few degrees of freedom. The application to fluid dynamics or, to be more precise, to spatially discretized Navier-Stokes flow, is far from straightforward.
In this historical review chapter, we follow the development of this field of research from the valiant naivety of the early 1980s to the open challenges of today.  
\end{abstract}
\maketitle

\section{Introduction}

In the nineteen seventies, there were two main modes of fluid dynamics research. This is aptly demonstrated by
T. B. Benjamin's influential papers on instabilities in viscous fluids, the first of which is titled
``Part I: theory'' and the second ``Part II: experiment''\cite{Benjamin1}-\cite{Benjamin2}. In this work, Benjamin emphasizes the importance of 
``a qualitative account of the exact problem''. By {\em the exact problem}, he referred to the necessity to retain the essential complications,
such as domain geometry and boundary conditions, in the mathematical model.
By {\em quantitative account}, he referred to a description of the branches of solutions and 
the way these are tied together at bifurcation points as control parameters are varied. The goal of this description is to predict
not the precise behaviour of the flow for particular initial conditions and parameter values, but rather the qualitative behaviour for any initial conditions and
a range of parameter values. A part of the bifurcation diagram can often be constructed using linear stability analysis and the theory of normal forms,
taking into account the spatial symmetries of the flow. This approach predicts what the diagram looks like locally, near
bifurcations of the base flow, thus providing pieces of a bigger puzzle. Other pieces can be provided by asymptotic expansions and careful experimentation.
Early examples of this approach can be found in Benjamin's collaborative work with Tom Mullin on ``Anomalous modes in the Taylor experiment'' \cite{Benjamin3} and
Busse and Clever's work on ``Instabilities of convection rolls in a fluid of moderate Prandtl number'' \cite{Busse}. While very successful, the qualitative approach
left a large part of the puzzle unsolved, restricted as it was to regimes with weakly nonlinear interaction and solutions sufficiently stable to
observe in experiments. In flows with strong nonlinearity and instability numerical techniques must be used -- an approach now considered a
third modus of fluid dynamics research, on par with theory and experiment.

\section{Pre-history}

Benjamin published the two parts mentioned above in 1978, when the numerical resolution of three-dimensional turbulent fluids was still in its
infancy. Only a few years earlier, Orszag and Patterson had dispelled ``much pessimism concerning the prospects for numerical simulation of 
three--dimensional turbulent flows'' \cite{OrsPat} by reporting on weakly turbulent flow in a periodic box, and the ground breaking work
of Kim, Moin and Moser\cite{kmm} on channel flow was still in progress. Similarly, techniques for the numerical computation of bifurcation diagrams were just
starting to be formulated and implemented. Doedel's work on AUTO\cite{Doedel} was likely the first attempt to create a software
package that enables the construction of a part of the diagram for a system specified by the user, thereby implementing many of the algorithms presented 
by Allgower and Georg \cite{AllGeo}. With the early versions of AUTO, one could approximate equilibria and periodic orbits in autonomous system of Ordinary Differential Equations (ODEs)
and locate their bifurcations. The computations did not depend on the availability of any kind of analytical expressions. Instead,
the equilibria or periodic orbits of interest are found as the solutions to well-posed Boundary Value Problems (BVP) through the use
of Newton-like iterative methods that are largely insensitive to dynamical instability and nonlinearity. The Lorenz '63 model \cite{lor63}, known in
the realm of fluid dynamics as a model for Rayleigh-B\'enard convection near its primary instability, was soon included as one of the demonstrations. The pitchfork bifurcation representing the onset of steady convection, and cascades of period doubling bifurcations at the onset of unsteady convection could readily be computed far away from parameter regimes accessible by asymptotic expansion. 

\section{The essential difficulty}

With simulation codes for discretized Navier-Stokes flow becoming available, and basic algorithms of numerical dynamical systems analysis implemented, the stage for
a numerical approach to a quantitative account of fluid dynamics was thus set in the early nineteen eighties. It would nevertheless take another twenty years before the stage
would fill with actors. A close look at the computational demands of this novel approach will explain the delay. Let us start with the comparatively simple task of
computing a branch of equilibria that bifurcates off an analytically known base flow. We will denote the system of ODEs that results from the spatial discretization of the flow by
\begin{equation}
\dot{\bm{x}}=\bm{f}(\bm{x},\alpha)
\label{NS_disc}
\end{equation}
where $\bm{x}$ is a vector of $n$ variables, the nature of which depends on the discretization, and $\alpha$ is a control parameter like the Reynolds number or an aspect ratio. 
We can consider the elements of $\bm{x}$ to be grid point values or expansion coefficients
with respect to some orthogonal basis of the fluid velocity or vorticity. We say that $\bm{x}$ is an element of the {\em phase space}, which is usually $\mathbb{R}^n$.
For simplicity, we will assume that the underlying grid, or truncation level, is equal for each spatial direction, so that
$n=\mbox{O}(N^3)$ for some $N$ that specifies the resolution. If we denote the base flow by $\bar{\bm{x}}$, its stability is determined by the eigenvalue problem 
\begin{equation}
D\bm{f}(\bar{\bm{x}},\alpha)\bm{v}=\lambda \bm{v}
\label{eigen}
\end{equation}
The Schur factorization of the $n\times n$ Jacobian matrix $D\bm{f}$ can be completed using a single $n\times n$ array taking up $8n^2$ bytes of memory assuming standard double precision 
arithmetic. If we assume we are working on
a state-of-the-art machine in the mid-eighties, the maximal Random Access Memory (RAM) would be about 8Mb, which gives $n=1000$ as the maximal matrix size. That, in turn,
would allow for less than ten grid points in each spatial direction to resolve three scalar fields. This may be enough to capture very smooth eigenfunctions. Let us assume that at
some critical value of the control parameter, $\alpha_{\rm c}$, an eigenvalue crosses zero and a branch of equilibria is created. In order to compute it, we can use parameter
continuation,
starting from an initial guess constructed from the critical eigenvector, computed at some $\alpha_o$ close to $\alpha_{\rm c}$. Setting the initial guess to 
$\bm{x}_0^{(0)}=\bar{\bm{x}}+c \bm{v}$ for some {\em ad hoc} parameter $c$, 
the continuation proceeds in a double loop:
\begin{itemize}
\item For $i=0\ldots \, N_{\rm cont}$ do\hfill{continuation loop}
\begin{itemize}
   \item[1.] For $j=0\ldots \, N_{\rm NR}$ do\hfill{Newton-Raphson loop}
   \begin{itemize}
      \item[a.] $\bm{r}=-\bm{f}(\bm{x}_i^{(j)})$
      \item[b.] If $\|\bm{r}\|<\epsilon$ exit Newton-Raphson loop
      \item[c.] Solve $D\bm{f}(\bm{x}_i^{(j)},\alpha) \delta\bm{x}=\bm{r}$
      \item[d.] $\bm{x}_i^{(j+1)}=\bm{x}_i^{(j)}+\delta \bm{x}$
   \end{itemize}
   \item[2.] If converged:\hfill{prediction step}
   \begin{itemize}
      \item[a.] $\bm{x}_{i}\leftarrow \bm{x}_i^{(j)}$
      \item[b.] $\bm{x}_{i+1}^{(0)}\leftarrow \bm{x}_{i}+\Delta \bm{T}_{i}$; $\alpha_{i+1}\leftarrow \alpha_i+ \Delta$
   \end{itemize}
             Else: reduce $\Delta$, recompute $\bm{x}_{i}^{(0)}$ and retry step 1.
\end{itemize}
\end{itemize}
Here, $T_{i}$ is an approximation of the tangent to the continuation curve, locally given by a function $\bm{x}(\alpha)$.
This is only a rudimentary outline of the  continuation algorithm -- a discussion of the appropriate value of the various parameters and the approximation of the tangent vector is beyond the scope of the current chapter. The important observation is that, in every iteration of the inner loop, we need $8n^2$ bytes of RAM to solve a linear system and this restricts the computations to a resolution that will likely prove insufficient after few continuation steps, as the secondary solution develops more spatial features. This consideration is only based on the memory requirements. In addition, we must consider the computation time, which is of order $\mbox{O}(n^3)=\mbox{O}(N^9)$ for the Schur decomposition and the linear solving, by default using LU-decomposition. This, too, posed a serious challenge at the time, the maximal processor speed being in the order of mega Herz. Assuming one had access to a high-end machine like the iPSC-1 (Intel Personal Super Computer) one could perform the Floating Point Operations (FLOPs) at a rate of about 25000 per second \cite{Bomans}. That would enable the completion of one Newton-Raphson iteration in about 7 hours. Assuming that, for a secondary solution close to its branch point, no more than four Newton-Raphson iterations are necessary for convergence, one could compute solutions at a rate of almost one per day. Even with hypothetical unlimited access to a supercomputer, the computation of a substantial part of the bifurcation diagram would be the work of months if not years. 

Of course, these estimates are overly simplified and several improvements can usually be made from the outset. For instance, if we consider incompressible flow, we can usually eliminate
one component of the fluid velocity. A second common reduction in the number of DOF stems from symmetries. Navier-Stokes flow in simple geometries, like pipes of constant diameter and
channels, have various rotation, reflection and shift symmetries. Solutions invariant under one or more of these symmetries are contained in a lower dimensional subspace of phase space and we can exploit
this fact to bring down the dimension of the linear problem in step 1c. For example, suppose we wanted to investigate the conclusions of the Lorenz model using a more realistic representation of thermal
convection, thereby
adhering to Benjamin's point of view. We would
replace the free-slip boundary conditions used by Saltzman \cite{Saltz} by no-slip conditions and switch from a Fourier basis to a Chebyshev basis in the wall-normal direction. 
Our first task would then be to compute equilibrium solutions that represent steady convection just beyond the pitchfork bifurcation of the conductive state. Since the governing
equations are equivariant under shifts in the spanwise direction, we can assume the equilibria to be independent of that direction. In addition, we can impose a point reflection
symmetry about the centre of the domain. As a result, we need to solve for two scalar functions, a stream function and the temperature, that both have a reflection symmetry. We can then 
take up to 40 Fourier functions in the streamwise direction and 40 Chebyshev polynomials in the wall-normal direction and still end up solving a linear system smaller than $1000 \times 1000$.
While still somewhat marginal, this is certainly a lot more promising.
Reduction by incompressibility and symmetry are ideas used extensively in the early explorations of bifurcations in fluid dynamics. Without pretending to write a comprehensive history, we
will follow two strands that differ in their approach while sharing a common goal.
 
\section{ENTWIFE}

The first starts with Benjamin himself. When he presented his work on bifurcations in Taylor-Couette flow in Harwell Laboratory, otherwise known as the Atomic Energy Research Establishment, near his home
university of Oxford, Andrew Cliffe was in the audience. At that time, Cliffe was part of a team developing finite-element codes for fluid dynamics simulations. He realized that he could use his simulation codes to find numerical evidence for the qualitative predictions Benjamin and Mullin had made\cite{Riley,DSWeb1}. This was the beginning of a long and fruitful collaboration resulting in numerous publications, the first of which sharing the title of Benjamin and Mullins earlier work \cite{Benjamin3} but prepended by ``A numerical and experimental study of \ldots''\cite{CliffeMullin}. In this paper results of eigenproblem and continuation algorithms like those presented above are described for axisymmetric solutions with up to four circulation cells. From the numbers of elements employed in a two-dimensional cross-section of the domain, we can deduce that Cliffe must have solved linear systems with several thousands of unknowns. This probably means that he made liberal use of  Harwell's supercomputer, which was the first in the United Kingdom (UK). 
It would seem that the solution with five cells was just out of reach, the authors reporting that ``a considerable amount of work will be required to investigate this case properly.''

The code that Cliffe used for this project grew into a software package called ENTWIFE \cite{Cliffe}. Based on finite-element discretization of fluid dynamics, it allowed for the continuation of equilibria and the detection of steady state and Hopf bifurcations. One could consider ENTWIFE an attempt to offer the capabilities of AUTO for discretized Partial Differential Equations (PDEs). Indeed, at the 1988 Workshop on Path-Following Methods in Leeds, UK, the two were presented side-by-side \cite{BKS}. The capabilities of AUTO to follow equilibria as well as periodic solutions, locate their bifurcations and switch branches at bifurcation points were demonstrated on the Lorenz model, while results obtained with ENTWIFE included the location of steady-state bifurcations in a two-dimensional convection problem. Theses results neatly demonstrate the trade-off between the level of sophistication of the analysis and the level of realism of the model under consideration imposed by the finite computational resources and numerical algorithms available at the time.

\section{Channel flow}

Also present at the conference in Leeds was Masato Nagata, who presented work on Taylor-Couette flow in the narrow-gap limit. Nagata's approach was different from that of Cliffe. Instead of discretization by finite elements, he used spectral decompositions of the flow. Finite elements are suitable for covering complex geometries like pipes with expansions or backward-facing steps. In contrast, spectral decompositions are suitable only for simple geometries like channels and pipes of constant diameter. Since the rate of convergence of spectral decompositions of the fluid velocity is usually far superior to that of finite-elements discretization, the former is the method of choice when available. However, the efficient spatial discretization alone was not enough to enable the computation of three-dimensional equilibria. In addition, Nagata exploited the incompressibility condition and several discrete symmetries. As a result, he could compute equilibria on a grid of about $16^3$ points \cite{DSWeb2}.

The computation of equilibria in Taylor-Couette flow was not the actual goal of Nagata's exploits. The real aim was
to compute finite-amplitude solutions in plane Couette flow. This was a rather ambitious project since, in the latter geometry, the laminar flow remains linearly stable for all Reynolds numbers and aspect ratios. In fact, around 1980, when Nagata started his PhD research with Busse at UCLA, it was not known whether any nontrivial equilibrium solutions even existed in plane Couette flow. His approach was to embed plane Couette flow in a family of flows that included forms of forcing other than friction at the walls. This family was parameterized in such a way that a nontrivial equilibrium, computed in the extended system, could be continued to the pure Couette case. This approach is now called the homotopy approach and was also used by
Cliffe and Mullin \cite{CliffeMullin}, who parameterized the boundary conditions to track the anomalous modes. These modes are connected to the laminar flow for idealized boundary conditions, but not for realistic, non-slip conditions.  

While in the study of anomalous modes in Taylor-Couette flow, the right homotopy could be deduced from symmetry considerations, for plane Couette flow there was no obvious way forward. Initially, Nagata added differential heating and placed the channel at an inclination. In this configuration, the base flow has an inflection point and exhibits a symmetry-breaking bifurcation from which two-dimensional solutions, consisting of transverse rolls, can be continued. In a subsequent bifurcation, three-dimensional equilibria are created \cite{nagata83}. While this work formed the basis of his PhD thesis, Nagata did not manage to complete the homotopy to plane Couette flow. His second attempt involved forcing through rotation and was, ultimately, successful. Under certain assumptions, the narrow-gap limit of Taylor-Couette flow coincides with plane Couette flow rotating about the spanwise direction. Taylor-Couette flow admits steady streamwise vortices, and these bifurcate into three-dimensional steady states. Nagata managed to trace some of these to the limit of zero rotation. Initially, he presented it not as a result but as a conjecture that this homotopy would yield genuine solutions in plane Couette flow \cite{nagata88}. The paper in which the central claim is staked did not appear until 1990, after a protracted battle with the reviewers \cite{Nagata90}. The sticky point in this battle was, unsurprisingly, the resolution. State-of-the-art direct numerical simulations used much higher resolution since they do not require the storage and decomposition of $n\times n$ matrices. The {\em highest available} resolution often getting confused with the {\em least sufficient} resolution, Nagata's results must have looked suspicious. After several years of pushing the limits of the computation, Nagata reports in the resulting paper that he used two different supercomputers: an IBM3081 and a CRAY, both of the same generation as the iPSC-1 mentioned above. The largest number of unknowns handled was $589$ and a single Newton-Raphson iteration could be completed in under four minutes. The final result was nonetheless hard to obtain since the rotation rate, which was the homotopy parameter, would often increase along sections of the continuation curves. As a consequence, Nagata had to compute a great many points on several continuation curves to obtain his final result.

\section{The limits of direct solving}

Nagata's results spurred a lot of work in the ``dynamical systems approach to turbulence'', as this field is now labeled in the subject classification of the American Mathematical Society (AMS).
One early contribution came from Fabian Waleffe, who started to formulate his theory about the way turbulence is sustained in channel flow as a postdoctoral researcher in the early nineties. Using a spectral approach, more accurate but slower than the pseudo-spectral methods introduced by Orszag and Patterson \cite{OrsPat}, he performed a series of numerical experiments to find the smallest possible computational domain that would sustain turbulence. In the course of these experiments, he observed that the fluid would occasionally approach travelling wave or time-periodic solutions. Using a Newton-Raphson scheme like the one described above, he managed to compute travelling wave solutions in Poiseuille-like flow and steady states in Couette-like flow \cite{wally98}. We use the suffix {\em -like} since Waleffe had made a compromise when formulating the model, prescribing stress instead of velocity at the boundaries. This allowed him to use a Fourier decomposition in all directions, resulting in a clean representation of the triad interactions of the Navier-Stokes equation and the conservation of energy in any truncated model. Later, he showed that a homotopy can be constructed to change the boundary conditions to non-slip without changing the qualitative features of the solutions\footnote{Some remarks can be found in the 2011 Woods Hole Lecture Notes \cite{WHOI}. See also later work by Chantry {\sl et al.} \cite{Chantry} on the validity of Waleffe's flow setup as a model for the inner region of channel flow.}. The analysis of the spatial structure of these invariant solutions cemented his theory about the regeneration cycle of large-scale streamwise rolls that was later investigated in pipe flow as well.
He ran his computations on grids of up to $8\times 20\times 8$ points (streamwise $\times$ wall-normal $\times$ spanwise), handling 3528 DOF. This corresponds to a 100Mb array to store the Jacobian matrix but, by this time, workstations were on the market with several hundreds of mega bytes of RAM, like the Sun SPARCSTATION available to Waleffe at MIT. He considered coding Newton-Raphson iteration for periodic solutions, resolving the time dependence with a Fourier series. However, this was still beyond the available RAM resources.

The first researchers to compute time-periodic solutions in the channel geometry were Genta Kawahara and Shigeo Kida, the former working in the group Nagata established at Kyoto 
University upon his return to Japan. Kawahara and Kida did not make modelling compromises, using a Fourier-Chebyshev basis and no-slip boundary conditions for plane Couette flow. The grid they used
had $16\times 31\times 16$ points and the flow was described by $15,422$ degrees of freedom. Rather than resolving the time dependence by Fourier series, they opted for a {\em shooting} approach, 
in which the periodic solution is represented by a single point on the orbit and the period. The defining system of equations is then
\begin{align}
\phi(\bm{x},P,\alpha)-\bm{x}&=0\\
\psi(\bm{x})&=0
\label{PO_BVP}
\end{align}
where $\phi$ is the solution to system (\ref{NS_disc}), $P$ is the period of the orbit and $\psi$ is a phase condition like a Poincar\'e plane of intersection. The Jacobian of this system, which would appear in the Newton-Raphson loop, is $(D\phi-\mathbb{I})$. However, since they did not have a machine with 2Gb of RAM at their disposal,
Kawahara and Kida had to strike a compromise with the method. Rather than Newton-Raphson iteration, they used a direction-set method that does not require derivatives of the objective function which, in this case, is the residual of the BVP, $\|\phi-\bm{x}\|+|\psi(\bm{x})|$. Evaluation of this function
requires only the simulation of the flow over one period.
The downside was that direction-set methods do not converge quadratically like Newton-Raphson iteration. The computations
ran day and night for many months on desktop computers. In the end, the results were accepted for publication when the relative residual, $\|\phi-\bm{x}\|/\|\bm{x}\|$, was less than $0.01$ \cite{KawKida01}.
 As the title of the resulting paper suggests, the emphasis was on the interpretation of the invariant solutions in terms of the cycle of regeneration of turbulence formulated by Waleffe.

A third line of research that deserves a separate mention is that on pipe flow. While the cylindrical geometry presents some technical challenges as compared to the rectangular geometry in channels, pipe flow is known to have qualitative properties similar to those of Couette and Poiseuille flow. In particular, the laminar Hagen-Poiseuille flow is known to be linearly stable at Reynolds numbers for which transient turbulence was observed, both in laboratory and in numerical experiments. Two groups simultaneously attempted to follow in Nagata's footsteps by computing finite-amplitude, fully nonlinear invariant solutions coexisting with stable laminar flow or, from a different perspective, to extend Waleffe's picture of the regeneration cycle to a different geometry. Their results were published around the same time, Faisst \& Eckhardt reporting on travelling waves with twofold and threefold rotational symmetry in a brief Letters paper and Wedin \& Kerswell presenting additional solutions with up to sixfold rotational symmetry in the Journal of Fluid Mechanics \cite{faisst03,wedin04}. Faisst \& Eckhardt used up to 5600 DOF, slightly exceeding Waleffe's largest computations, while Wedin \& Kerswell pushed the limits by considering 20,000 DOF, thereby consuming up to 3Gb of RAM. It was almost certainly the most memory-intense computation of invariant solutions ever performed, and required the purchase of a brand new DEC Alpha computer with 4GB of RAM and two CPUs. While the DEC Alpha presented a staggering ten thousand fold increase in performance and five hundred fold increase in RAM with respect to the best available machines of the early eighties, the demands posed by the most daring new computations soon outstripped the the latest technology. New computational algorithms were necessary to bridge the gap.

\section{Experimental and technical break-through}
While the results that appeared in the eighties and nineties served as a proof of principle and demonstrated the potential of the dynamical systems approach to shine  new light on outstanding
questions in fluid dynamics, they
were all severely restricted by memory and computation time requirements. It took two breakthroughs for the research in this area to really take off. One was the experimental observation
of travelling waves in pipe flow by Bj\"orn Hof and coworkers \cite{Hofetal} and the second was the
widespread adoption of Krylov subspace methods. Both took place in the early 2000s.
When one considers the citation record of the seminal works by Nagata \cite{Nagata90}, Waleffe \cite{wally98} and Kawahara and Kida \cite{KawKida01} combined, a fourfold increase of the citation rate shows
around 2004, as illustrated in figure \ref{papersandcitations}.

One of the reasons why the early work on channel flow did not immediately get much traction was likely that the results could be verified neither by analysis nor by experiment. For many a fluid physicist,
considering Navier-Stokes flow in a perfectly regular channel as a mathematical idealization in the first place,
the artificial systems considered in the homotopy method, the low resolutions and Reynolds numbers and the small computational domains must have felt alien and unconvincing. While work on symmetry-breaking transitions from a laminar
base flow could be verified by experiments like those of Mullin, the computation of finite-amplitude, fully nonlinear equilibrium solutions, co-existing with a stable base flow, seemed
impossible to relate to measurement. The team lead by Hof, however, managed to do exactly that. They measured transient velocity profiles in a pipe and compared them with computed travelling wave
solutions. Even though the Reynolds number differs between the experiment and the computations, the agreement was convincing. In particular, the discrete rotational symmetry
that holds exactly for the numerically computed solution holds approximately for the experimental data. This work can be considered the first proof of principle that results of the dynamical systems approach actually had a bearing on open question in fluid physics far from near-laminar or weakly nonlinear regimes. In later work by a partially overlapping team, the link between experiment and computation was pushed to the point where travelling wave solutions could be computed by Newton-like iterative methods, starting from appropriately smoothed experimental data \cite{Lozar}.

The adoption of Krylov subspace methods presented a technical break-through. These so-called {\em matrix-free} methods can be used to solve the eigenproblems like (\ref{eigen}) and linear systems like that in step 1c. of the continuation algorithm. As the name suggests, Krylov subspace methods avoid the computation and decomposition of any $n\times n$ matrix and thereby eliminate the memory problem. In particular, let us consider the Generalized Minimal RESidual (GMRES) method which, in its simplest form, works like this:
\begin{itemize}
\item[1.] Set $\bm{v}_0=\bm{b}$
\item[2.] For $j=1\ldots N_{\rm K}$ do \hfill{Krylov loop}
\begin{itemize}
   \item[a.] Let $\bm{v}_j=\bm{A}\bm{v}_{j-1}$
   \item[b.] Find the vector $\bm{x}^*\in \text{Span}\{\bm{v}_0,\ldots,\bm{v}_j\}$ that minimizes the residual $\|\bm{A}\bm{x}-\bm{b}\|$
   \item[c.] If $\|\bm{A}\bm{x}^*-\bm{b}\|<\epsilon$ exit Krylov loop
\end{itemize}
\item[3.] If converged output $\bm{x}^*$ as approximate solution to $\bm{A}\bm{x}=\bm{b}$.
\end{itemize}
In words, the solution to the linear problem is approximated after $j$ steps of the algorithm by minimizing the residual over the $(j+1)$-dimensional Krylov subspace $\text{Span}\{\bm{v}_0,\ldots,\bm{v}_j\}$. 
The algorithm shown here is practical only for well-conditioned matrices. If the singular values of $\bm{A}$ vary greatly in magnitude, as they usually will in applications to Navier-Stokes flow, the Krylov basis $\text{Span}\{\bm{v}_0,\ldots,\bm{v}_j\}$ quickly becomes nearly degenerate and the minimizer cannot be found with a reasonable degree if accuracy. To circumvent this problem, a full-fledged implementation of GMRES includes an intermediate step in which the basis is made orthogonal. A nearly optimal way to do this is described by Saad and Schultz \cite{SS}. The original work by A. N. Krylov does not cover this important step, which is one reason why it is not commonly cited as opposed to the work by Saad and Schultz. Another reason may be that Krylov's paper is only available in Russian \cite{Krylov}. 
In order to appreciate the efficiency of this matrix-free solver as compared to direct methods, let us briefly look into its convergence properties.
GMRES converges well if the following conditions hold:
\begin{enumerate}
\item the eigenvalues of $\bm{A}$ are clustered,
\item the eigenvalues of $\bm{A}$ are bounded away from zero and
\item $\bm{A}$ is not too far from normal.
\end{enumerate}
The linear problem in step 1c. of the continuation algorithm does not normally satisfy these conditions. There, $\bm{A}=D\bm{f}$ is the Jacobian of the discretized Navier-Stokes equation. Since this equation includes the Laplacian, which is an unbounded operator, its eigenvalues will not be clustered but rather extend to infinity. One possible solution is to pre-condition the linear problem, for instance by solving instead the system $\bm{B}\bm{A}\bm{x}=\bm{B}\bm{b}$ for some matrix $\bm{B}$ chosen such that the product $\bm{B}\bm{A}\bm{v}$ can easily be computed for any $\bm{v}$ and $\bm{B}\bm{A}$ has clustered eigenvalues. This strategy had been explored by Tuckerman for the computation of equilibria in cylindrical Rayleigh-B\'enard convection as early as 1989 -- while Nagata was still battling reviewers over the spatial resolution of his now classical results \cite{tuckerman2}. In this short paper, she showed that pre-conditioning by the inverse of the Laplacian can be achieved by re-purposing the code for a first-order semi-implicit time stepper. This method, known as {\em Stokes preconditioning}, may have been formulated ahead of its time and did not immediately get a lot of traction beyond Tuckerman's academic descendents. However, its use increased greatly in the 2000s and a plethora of results based on this approach have now been published on a variety of flows, most notably on doubly diffucive convection (see, e.g. \cite{BK}).

An alternative solution, proposed by S\'anchez {\sl et al.} \cite{sanch2}, is to reformulate the BVP to take the same shape as that for time-periodic solutions (\ref{PO_BVP}):
\begin{align}
\phi(\bm{x},P,\alpha)-\bm{x}&=0\\
P-c&=0
\label{EQ_BVP}
\end{align}
In words, we find the equilibrium as the fixed point of the flow over an arbitrary time $c$. The matrix in the linear system then is $(D\phi-\mathbb{I})$. we can compute the action of this matrix on a perturbation vector by integrating the discretized equations (\ref{NS_disc}) along with the tangent linear model
\begin{equation}
\dot{\bm{v}}=D\bm{F}(\bm{x})\bm{v}
\label{NS_disc_pert}
\end{equation}
over time $c$, starting from the current approximate equilibrium point and the perturbation provided by GMRES.
This reformulation is beneficial because the eigenvalues $\lambda_i$ of $D\bm{F}$ are related to those of 
the matrix of derivatives of the flow, $D\phi$, through the exponential map $\mu_i=\exp(c\lambda_i)$. Thus, all eigenvalues $\lambda_i$ with large, negative real part, related to the Laplacian operator,
correspond to eigenvalues in a small disc around minus unity of the matrix $(D\phi-\mathbb{I})$. The larger we make the parameter $c$, the more clustered these eigenvalues become. At the same time, the computation time of each GMRES iterations increases with $c$. Choosing the optimal value for the integration time is a balancing act. The Jacobian matrix associated with the BVP for periodic solutions has a clustered spectrum for the same reason, but in that case the integration time is set by the dynamics of the flow\footnote{This presentation is, in fact, an anachronism. The original paper by S\'anchez {\sl et al.} \cite{sanch2} focuses on periodic solutions. The unified presentation, with the phase condition differentiating between periodic and equilibrium solutions, can be found in a recent review by S\'anchez and Net \cite{SN}.}. In summary, both the BVP for equilibria (\ref{EQ_BVP}) and for periodic orbits (\ref{PO_BVP}) give rise to linear systems that are naturally suitable for GMRES.

For either strategy, the question remains how large $N_{\rm K}$ must be in order to find an accurate enough Newton-Raphson update step to maintain convergence. An quantitative answer to this question will depend on the details of the system under consideration, but we can give a qualitative view. The eigenvalues that become strongly clustered are those related to the Laplacian operator, i.e. the viscous term in the Navier-Stokes equation. The associated eigenmodes have a fine spatial scale. One can think of the Kolmogorov length scale as separating the eigenmodes dominated by viscous damping from those dominated by nonlinear processes. It seems reasonable to assume, then, that the number of GMRES iterations needed will grow algebraically with the Reynolds number, for instance as the classical estimate $Re^{9/4}$ of the number of active modes in turbulent flow by Constantin {\it et al.} \cite{CFMT}. As a consequence, the Newton-Krylov iteration described here may not work well at high Reynolds numbers, and we will discuss some open challenges below. For flows close to the onset of turbulence, excellent results have been reported. A review of results obtained with Stokes preconditioning was presented by Tuckerman {\sl et al.} \cite{TLW}. This paper also contains a comprison between the two strategies for computing equilibria which demonstrates the Stokes preconditioning can be significantly more efficient than ``preconditioning by time-stepping''. Results obtained with the latter method include work by
  S\'anchez {\it et al.} \cite{sanch2}, who computed periodic solutions in annular thermal convection, truncating the discretization to 31,870 DOF and using fewer than 60 GMRES iterations. Thus, they achieved a reduction of a factor of over 500 of RAM usage and computation time as compared to direct solving. Soon afterwards, Viswanath computed a number of periodic solutions in plane Couette flow with 319,790 DOF \cite{viswanath07}. He also presented a careful analysis of the truncation error, stressing that increasing the resolution so drastically over that used by Kawahara \& Kida was not a vanity project, but actually necessary to accurately resolve the fluid motion.

\section{Logistic growth}

\begin{figure}
\begin{center}
\includegraphics[width=0.6\textwidth]{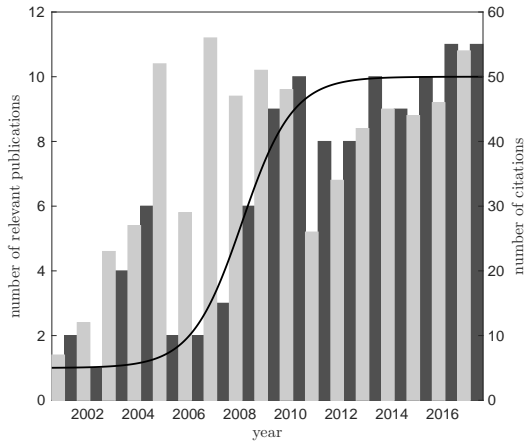}
\end{center}
\caption{Light grey bars: the number of citations per year of the three seminal papers by Nagata \cite{Nagata90}, Waleffe \cite{wally98} and Kawahara and Kida \cite{KawKida01} combined. Dark grey bars: an estimate of the number of papers published on the analysis of fluid dynamics with computational dynamical systems techniques per year since 2000. Both show a sharp increase after the appearance of the landmark papers by Hof {\sl et al.} \cite{Hofetal} and S\'anchez {\sl et al.} \cite{sanch2} in 2004, the latter with a few years delay as Newton-Krylov codes were being developed in various research groups. These data were obtained from the Web of Science by searching for key words such as ``exact coherent structures'' and ``travelling waves'', as well as by examining citations to highly cited papers. Relevance of the counted papers was established by the author, and the numbers should be interpreted as estimates only. The solid line somewhat speculatively denotes logistic growth.}
\label{papersandcitations}
\end{figure}

The ``triple loop'' approach, with the GMRES loop inserted at the place of the linear solve in step 1c. of the continuation algorithm, quickly became the standard. With that, the resolution of the underlying fluid simulation was no longer bound by memory restrictions. At the same time, computers got faster and parallel processing became widely available, so that simulations ran fast even at resolutions similar to those used by Viswanath. The combination of matrix-free methods with parallel processing promised an unprecedented insight in the spatio-temporal structure of turbulence at low to moderate Reynolds numbers.
In the decade that followed, too many results followed to summarize here. Many concerned the relatively simple geometry of channels, pipes and ducts. Some were obtained with legacy time-stepping codes, coupled to newly written Newton-Krylov routines, while others used new software written especially for the purpose of dynamical-systems type computations, such as Gibson's Channelflow code \cite{channelflow} and Willis' Openpipeflow \cite{openpipeflow}. 
Overviews of such work can be found in Annual Review papers by Eckhardt \cite{eckhardt07} and Kawahara {\it et al.} \cite{KUV}. Work in a different vein was performed with the LOCA code {\sl et al.} \cite{Salinger}. In a sense, the latter code took the baton from ENTWIFE, employing finite elements and sophisticated meshing and preconditioning techniques. Indeed, the main author, Andrew Salinger, published several papers with Andrew Cliffe. The use of finite elements allowed them to tackle more complicated geometries, such as that of two colliding jets in a annular domain \cite{pawlowski06}.

The number of researchers active in the field grew, and various workshops dedicated to the dynamical systems approach were held, such as in Bristol in May 2004, at the Newton Institute in September 2008, in Carg\`ese in France in May 2014 and in Santa Barbara in January 2017 \cite{meetings}. Dedicated
sessions became a regular fixture at the Conference on Applications of Dynamical Systems of the Society for Industrial and Applied Mathematics (SIAM) and annual meeting of the Division of Fluid Mechanics of the American Physical Society (APS). In fact, if one wants to, one can see an exponential increase in the amount
of published work in the period from 2004 to 2009. After that, the rate of production of novel results leveled off. The comparatively low-hanging fruit had been plucked, and more challenging problems begged investigating -- problems that required the development of novel algorithms, the optimization and parallelization of existing codes and similar major investments of time and energy. Some problems seemed more amenable to an approach rooted in statistical physics instead of dynamical systems theory, and this work drifted out of the scope of the current narrative accordingly.
The initial rapid increase and subsequent levelling gives the publication statistics over the period from 2004 to the present day the shape of logistic growth, as illustrated in figure \ref{papersandcitations}. 

\section{Conclusion}

What started with visionary ideas that the computers of the nineteen eighties could not quite support has evolved into a lively and mature field of research. While the days of exponential growth may be over, new directions are still being explored. To mention but a few recent developments, we saw the investigation of the asymptotic suction boundary layer by Khapko {\sl et al.} \cite{Khapko}, beautiful experiments on quasi-two dimensional Kolmogorov flow by Suri {\sl et al.} \cite{suri} and the application of dynamical systems computations to large eddy simulation by Hwang {\sl et al.} \cite{Hwang}, Sasaki {\sl et al.} \cite{Sasaki} and Sekimoto and Jim\'enez \cite{Sekimoto}. There are many more avenues to pursue, and I hope that the current historical overview will not only entertain, but also help put new endeavours in perspective and perhaps give some inspiration. 

\section*{Acknowledgements}

I would like to thank Sebastian Altmeyer, Andrew Hazel, Bj\"orn Hof, Genta Kawahara, Rich Kerswell, Masato Nagata, Laurette Tuckerman and Fabian Waleffe for sharing their ideas, memories and corrections.

\bibliographystyle{spbasic}
\bibliography{A_brief_history}
\end{document}